\documentclass[onecolumn,12pt,draftcls]{IEEEtran}

\IEEEoverridecommandlockouts                              
\usepackage{epsfig} 
\usepackage{amsmath} 
\usepackage{amssymb} 

\title{\LARGE \bf
Coding  for Additive White Noise Channels with Feedback Corrupted by
Uniform Quantization or Bounded Noise }

\author{ Nuno C Martins and Tsachy Weissman
\thanks{({\tt \small nmartins@umd.edu}) Nuno C. Martins is with the Electrical and Computer Engineering Department and the Institute for Systems Research at the University of Maryland, College Park. ({\tt \small tsachy@stanford.edu}) Tsachy Weissman is with the Department of Electrical Engineering and the Information Systems Laboratory at Stanford University. Note: An abridged version of this work
was presented at Stanford University on July 7th of 2006, in the Colloquium on Feedback Communications. }%
}

\begin{document}

\maketitle
\thispagestyle{empty}
\pagestyle{empty}

\newtheorem{theorem}{Theorem}[section]
\newtheorem{corollary}[theorem]{Corollary}
\newtheorem{conjecture}[theorem]{Conjecture}
\newtheorem{lemma}[theorem]{Lemma}
\newtheorem{proposition}[theorem]{Proposition}
\newtheorem{axiom}{Axiom}[section]
\newtheorem{remark}{Remark}[section]
\newtheorem{example}{Example}[section]
\newtheorem{exercise}{Exercise}[section]
\newtheorem{definition}{Definition}[section]
\newtheorem{theorema}{Theorem A}
\newtheorem{lemmaa}{Lemma A}
\newtheorem{remarka}{Remark A}
\newtheorem{corollarya}[theorem]{Corollary A}

\begin{abstract}
We present simple coding strategies, which are variants of the
Schalkwijk-Kailath scheme, for communicating reliably over additive
white noise channels in the presence of corrupted feedback. More specifically, we consider a framework comprising an additive white forward channel and a backward link which is used for feedback. 
We consider two types of corruption mechanisms in the backward link. The first is quantization noise, i.e., the encoder receives the quantized values of the past outputs of the forward channel. The quantization is uniform, memoryless and time invariant (that is, symbol-by-symbol scalar quantization), with bounded quantization error. The second corruption mechanism is an arbitrarily distributed additive bounded noise in the backward link. Here we allow symbol-by-symbol encoding at the input to the backward channel. We propose simple explicit schemes that guarantee positive information rate, in bits per channel use, with positive error exponent.
If the forward channel is additive white Gaussian then our schemes achieve capacity, in the
limit of diminishing amplitude of the noise components at the
backward link, while guaranteeing that the probability of error converges to zero as a doubly exponential function of the block length. 
Furthermore, if the forward channel is additive white Gaussian and the backward link consists of an
 additive bounded noise channel, with signal-to-noise ratio (SNR) constrained symbol-by-symbol encoding, then our schemes
are also capacity-achieving in the limit of high SNR. 
\end{abstract}

\section{Introduction}
\label{sec:Introduction}

That noiseless  feedback does not increase the capacity of
memoryless channels, but can dramatically enhance the reliability
and  simplicity of the schemes that achieve it, is well known since
Shannon's  work \cite{Shannon56}. The assumption of noiseless
feedback is an idealization often meant to capture communication
scenarios where the noise in the backward link is significantly
smaller than in the forward channel. However, all the known simple
schemes for reliable communication in the presence of feedback rely
heavily on the assumption that the feedback is completely
noise-free, and break down when noise is introduced into the
backward link.

As a case in point, it was recently shown in \cite{KimLapidothWeissman}
that \emph{any} feedback scheme with linear encoding (of which the
Schalkwijk-Kailath scheme and its variants are special cases) breaks
down completely in the presence of additive white noise of
arbitrarily small variance in the backward link: not only is it
impossible to achieve capacity, but, with such schemes it is
impossible to communicate reliably at any positive information rate.

It is therefore of primary importance, from both the theoretical and
the practical viewpoints, to develop channel coding schemes that, by
making use of \emph{noisy} feedback, maintain the simplicity of
noiseless feedback schemes while achieving a positive rate of
 reliable communication. It is the quest for such schemes that
 motivates this paper.

Our main contribution is the derivation of simple coding strategies,
which are variants of the Schalkwijk-Kailath scheme, for
communicating over additive white channels in the presence of corrupted
feedback. More specifically, we consider
 two  types of corruption mechanisms in the backward link:
  \begin{itemize}
    \item Quantization noise: the encoder receives the quantized
    values of the past outputs of the forward channel. The quantization is
    uniform, memoryless and time invariant (that is, symbol-by-symbol scalar
    quantization), with  bounded quantization error.

    \item Additive bounded noise:  the noise in the backward
    link is additive, and has bounded components, but is otherwise
    arbitrarily distributed. Here we allow symbol-by-symbol
    encoding at the input to the backward channel.

  \end{itemize}
The coding schemes that we present achieve positive information rate with positive error exponent. In addition, 
if the forward channel is additive white Gaussian then our schemes are
capacity-achieving, in the limit of diminishing amplitude of the
noise components in the backward link. Furthermore, if the backward link consists of an additive bounded noise channel, with instantaneous encoding, then our schemes
are also capacity-achieving in the limit of high SNR (in the backward
link). We note that the diminishing of the gap to capacity with
vanishing noise in the backward link is a desired property, not to
be taken for granted in light of the negative results in
\cite{KimLapidothWeissman}. In addition, the probability of error of our coding schemes converges to zero as a doubly exponential function of the block length,
provided that the forward channel is additive, white and Gaussian. As will be seen in subsequent sections,
our analysis of the performance of the suggested schemes is  based
on elementary linear systems theory. 

To our knowledge, the impact of noise in the feedback link on
fundamental performance limits and on explicit schemes that attain them
has hitherto received little attention. Exceptions are the
papers \cite{SahaiSimsek, DraperSahai} which study the trade-off
between reliability and delay in coding for discrete memoryless channels with noisy
feedback,  and suggest concrete coding schemes for this scenario.
Another exception is the recent \cite{Permuter}, which considers
the capacity of discrete finite-state channels in the presence
 of non-invertible maps in the feedback link, such as  quantization.
Yet another paper is the aforementioned  \cite{KimLapidothWeissman},
which is primarily concerned with  the impact of noise in the
backward link on the error exponents. 

The remainder of this paper is structured as follows. Section
\ref{sec:726pm17jun06} presents preliminary results and definitions, while
Section \ref{sec:231pm05june06} specifies and analyzes a coding scheme
in the presence of feedback corrupted by bounded additive noise,
under the assumption that the noise is observable at the decoder.
The main results of the paper are presented in Sections
\ref{sec:447pm15jun06} and \ref{sec:639pm17jun06}, where we describe and analyze
coding schemes for the cases where the backward link features uniform quantization or bounded additive noise, respectively.
The paper ends with conclusions in Section \ref{sec:conclusions}.

\textbf{Notation:}
\begin{itemize}
\item Random variables are represented in large caps, such as $Z$. 
\item Stochastic processes are indexed by the discrete time variable $t$, like in $X_t$. We also use $X^t$ to represent $(X_0, \ldots, X_t)$,
provided that $t \geq 0$. If $t$ is a negative integer then we adopt the convention that $X^t$ is the empty set. 
\item A realization of a random variable $Z$ is represented in small caps, such as $z$.
\end{itemize}

\section{Preliminary Results and Definitions}
\label{sec:726pm17jun06}
\begin{figure}
\centering

\epsfig{file=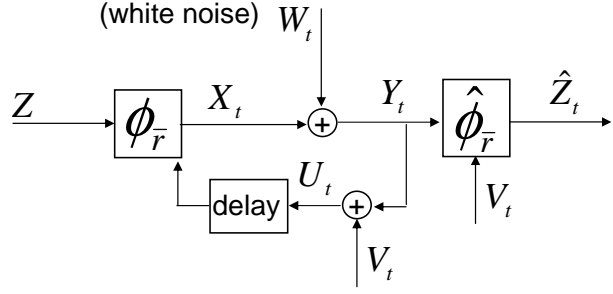,height=1.5 in,angle=0}

\caption{Basic feedback scheme.}
\label{Fig:904pm30may06}
\end{figure}

In this Section, we define and analyze  a feedback system
whose structure is described by the diagram of Fig \ref{Fig:904pm30may06}. The aforementioned
system will be present in the coding schemes proposed in subsequent Sections.

For the remainder of this paper, we consider that $W_t$ is a
zero mean and white stochastic process of
variance $\sigma_W^2$ and that $Z$ is a real random variable taking
values in
$[0,1]$. In addition, $Z$ and $W^t$ are assumed independent for all
$t$. The feedback noise $V_t$ is a bounded real stochastic process 
whose amplitude has a least upper-bound given by:
$$\bar{\sigma}_V \overset{def}{=} \inf \{ \alpha \in \mathbb{R}_{\geq 0} : Prob(|V_t|>\alpha)=0, t \geq 0\}
$$ meaning that the following holds: $$Prob(|V_t| \leq \bar{\sigma}_V)=1, \text{ } t \geq 0$$ 
The remaining signals $U_t$, $Y_t$ and $\hat{Z}_t$ are
also real stochastic processes. The block represented in Fig
\ref{Fig:904pm30may06} by $\phi_{\bar{r}}$ is an operator that maps
$Z$ and $U^{t-1}$ into
$X_t$ for all $t$. Similarly, $\hat{\phi}_{\bar{r}}$ maps $Y^t$ and
$V^t$ into $\hat{Z}_t$. The description of the maps $\phi_{\bar{r}}$ and $\hat{\phi}_{\bar{r}}$ is given in the following definition.

\begin{definition}
\label{def:1011pm30may06}
 Given a positive real constant $\bar{r}$, the operators $\phi_{\bar{r}}: \left( t,Z,U^{t-1} \right) \mapsto X_t$ and $\hat{\phi}_{\bar{r}}: \left( t,Y^t,V^{t} \right) \mapsto \hat{Z}_t$, represented in Fig \ref{Fig:904pm30may06},
 are defined as follows:
\begin{equation}
\label{eq:701pm31may06}
X_{t}=\phi_{\bar{r}}\left(t, Z,U^{t-1} \right) \overset{def}{=} \begin{cases} (2^{-\bar{r}}-2^{\bar{r}}) \left( \sum_{i=0}^{t-1} 2^{\bar{r}(t-i-1)}U_i+2^{\bar{r}t}Z \right) & \text{if $t \geq 1$} \\ (2^{-\bar{r}}-2^{\bar{r}})Z & \text{if $t=0$} \end{cases}
\end{equation}
\begin{equation}
\label{eq:702pm31may06}
\hat{Z}_{t}= \hat{\phi}_{\bar{r}}\left(t, Y^t,V^t \right) \overset{def}{=} \begin{cases} -\sum_{i=0}^{t-1} 2^{-\bar{r}(i+1)} (V_{i}+Y_{i})  & \text{ if $t \geq 1$} \\ 0 & \text{if $t=0$} \end{cases}
\end{equation}

\end{definition}

Notice that (\ref{eq:701pm31may06}) has a term, given by $2^{\bar{r}t}Z$, that grows exponentially. However, it should be observed that if the feedback loop is closed (see Fig \ref{Fig:904pm30may06})
by using $U_t=X_t+V_t+W_t$ then $X_t$ is given by:
\begin{equation}
\label{eq:12368sep06}
X_t=(2^{-\bar{r}}-2^{\bar{r}})\left( \sum_{i=0}^{t-1} 2^{-\bar{r}(t-i-1)}(W_i+V_i)+2^{-\bar{r}t}Z\right), \text{ } t \geq 1
\end{equation} which describes a system that is stable, in the bounded input implies bounded output sense. In the absence of backward link noise, i.e. $V_t=0$, (\ref{eq:701pm31may06}) and (\ref{eq:702pm31may06}) are equivalent to the equations used in the original work by  Schalkwijk-Kailath \cite{SchalkwijkKailath}. An alternative minimum variance control interpretation to (\ref{eq:701pm31may06}) and (\ref{eq:702pm31may06}), in the presence of \emph{perfect} feedback, is given in \cite{Elia04}. In addition, the work by \cite{Elia04} extends Schalkwijk-Kailath's algorithm, with \emph{perfect} feedback, to the multi-user case. A general control theoretic framework to feedback capacity is given in \cite{Tatikonda}. The following lemma states a few properties of (\ref{eq:701pm31may06}) and (\ref{eq:702pm31may06}) which motivate their use in the construction of coding schemes.

\begin{lemma}
Let $\sigma_W^2$, $\bar{\sigma}_V$ and $\bar{r}$ be given positive real constants. Consider the feedback system of Fig \ref{Fig:904pm30may06}, which is described by (\ref{eq:701pm31may06})-(\ref{eq:702pm31may06}) in conjunction with the following equations:
\begin{equation}
Y_t=X_t+W_t
\end{equation}
\begin{equation}
U_t=X_t+V_t+W_t
\end{equation} The following holds:
\begin{equation}
\label{eq:932pm31may06}
 X_t=2^{\bar{r}t}(2^{\bar{r}}-2^{-\bar{r}})(\hat{Z}_t - Z), \text{ }  t \geq 0
\end{equation} 
\begin{equation}
\label{eq:1019pm30may06}
 E[X_t^2] \leq \left( \sigma_W \sqrt{2^{2\bar{r}}-1}+\bar{\sigma}_V (2^{\bar{r}}+1) + 2^{-\bar{r}t} (2^{\bar{r}}-2^{-\bar{r}}) \right)^2, \text{ }  t \geq 0
\end{equation}
If $W_t$ is zero-mean, white and Gaussian, with variance $\sigma_W^2$, then the following holds:
\begin{equation}
\label{eq:257pm4sept06}
 Prob \left( |X_t| \geq \alpha \right) \leq e^{-\frac{\left( \alpha-\gamma\right)^2}{2 \beta^2}}, \text{ } \alpha > 0,  \text{ } t \geq 0
\end{equation} where $\gamma$ and $\beta$ are the following positive real constants:

\begin{equation}
\label{eq:258pm4sept06}
 \gamma \overset{def}{=} \left( 2^{\bar{r}} + 1 \right) \bar{\sigma}_V + 2^{\bar{r}}-2^{-\bar{r}}
\end{equation}
\begin{equation}
\label{eq:259pm4sept06}
\beta^2 \overset{def}{=} \left( 2^{2\bar{r}}-1 \right) \sigma_W^2
\end{equation}
\end{lemma}

\textbf{Proof:}
In order to derive (\ref{eq:932pm31may06}), we substitute $U_t=V_t+Y_t$ in (\ref{eq:702pm31may06}).
We now proceed to proving the validity of (\ref{eq:1019pm30may06}). Since the operators $\phi_{\bar{r}}$ and $\hat{\phi}_{\bar{r}}$ are linear, we can bound the variance of $X_t$ by separately quantifying the contribution of the external \emph{inputs} $Z$, $W_t$ and $V_t$. By making use of the triangular inequality, we arrive at the following bound:
\begin{equation}
\label{eq:1036pm30may06}
\left( E[X_t^2] \right)^{1/2} \leq \left( \sigma_W^2 \frac{1}{2 \pi} \int_{-\pi}^{\pi} \left|T \left( e^{j\omega} \right) \right|^2 d \omega \right)^{1/2}+ \bar{\sigma}_V \max_{\omega \in (-\pi,\pi]} \left|T \left( e^{j\omega} \right) \right|+2^{-\bar{r}t}(2^{\bar{r}}-2^{-\bar{r}})
\end{equation}
where $T\left( e^{j\omega} \right)$ is the following transfer function:
\begin{equation}
T\left( e^{j\omega} \right) = \frac{2^{-\bar{r}}-2^{\bar{r}}}{e^{j\omega}-2^{-\bar{r}}}
\end{equation}
The transfer function $T\left( e^{j\omega} \right)$ describes the input-output behavior of the feedback loop from $V_t$ to $X_t$ and from $W_t$ to $X_t$. The first term in the right hand side of (\ref{eq:1036pm30may06}) quantifies the contribution from the white process $W_t$, while the second term is an upper-bound to the contribution of $V_t$ and the last term comes from the \emph{initial condition} determined by $Z$. Standard computations lead to the following results:
\begin{equation}
\label{eq:1044pm30may06}
\frac{1}{2 \pi} \int_{-\pi}^{\pi} \left|T \left( e^{j\omega} \right) \right|^2 d \omega = 2^{2\bar{r}}-1
\end{equation}
\begin{equation}
\label{eq:1045pm30may06}
\max_{\omega \in (-\pi,\pi]} \left|T \left( e^{j\omega} \right) \right| = \frac{2^{\bar{r}}-2^{-\bar{r}}}{1-2^{-\bar{r}}} = 2^{\bar{r}}+1
\end{equation}
After substituting (\ref{eq:1044pm30may06}) and  (\ref{eq:1045pm30may06}) in (\ref{eq:1036pm30may06}), we arrive at (\ref{eq:1019pm30may06}).
In order to prove (\ref{eq:257pm4sept06})-(\ref{eq:259pm4sept06}), under the assumption that $W_t$ is zero mean white Gaussian, we define the following auxiliary Gaussian process:
\begin{equation}
\tilde{X}_t= \begin{cases} 0 & \text{if $t=0$} \\ (2^{-\bar{r}}-2^{\bar{r}}) \sum_{i=0}^{t-1} 2^{-\bar{r}(t-i-1)} W_i & \text{if $t \geq 1$} \end{cases}
\end{equation}
After simple manipulations, similar to the ones leading to (\ref{eq:1044pm30may06})-(\ref{eq:1045pm30may06}), we get the following properties of $\tilde{X}_t$:
\begin{equation}
E[\tilde{X}_t^2] = \left( 2^{2\bar{r}}-1 \right) \left( 1-2^{-2\bar{r}t} \right) \sigma_W^2 \leq \beta^2
\end{equation}
\begin{equation}
|\tilde{X}_t -X_t| \leq \left( 2^{\bar{r}}-2^{-\bar{r}} \right) \left( \bar{\sigma}_V \frac{1-2^{-\bar{r}t}}{1-2^{-\bar{r}}}+2^{-\bar{r}t} \right) \leq \gamma
\end{equation} where we used the definitions (\ref{eq:258pm4sept06}) and (\ref{eq:259pm4sept06}) along with (\ref{eq:12368sep06}). Consequently, we arrive at:
\begin{equation}
Prob \left( |X_t|  \geq \alpha \right) \leq Prob \left( |\tilde{X}_t| \geq \alpha-\gamma \right) \leq \sqrt{\frac{2}{\pi \beta^2}} \int_{\alpha-\gamma}^{\infty} e^{-\frac{\mu^2}{2 \beta^2}} d \mu, \text{ } \alpha > 0
\end{equation} where we used the facts that, by definition, $|\tilde{X}_t -X_t| \leq \gamma$, that $E[\tilde{X}_t^2] \leq \beta^2$ and that $\tilde{X}_t$ is normally distributed.
The derivation of (\ref{eq:257pm4sept06}) is complete once we use the following upper-bound \cite[page 220 eq. (5.1.8)]{Proakis}:
\begin{equation}
\sqrt{\frac{2}{\pi \beta^2}} \int_{\alpha-\gamma}^{\infty} e^{-\frac{\mu^2}{2 \beta^2}} d \mu \leq  e^{-\frac{\left( \alpha-\gamma\right)^2}{2 \beta^2}}
\end{equation}
$\square$

\section{A coding scheme with feedback}
\label{sec:231pm05june06}

\begin{figure}
\centering

\epsfig{file=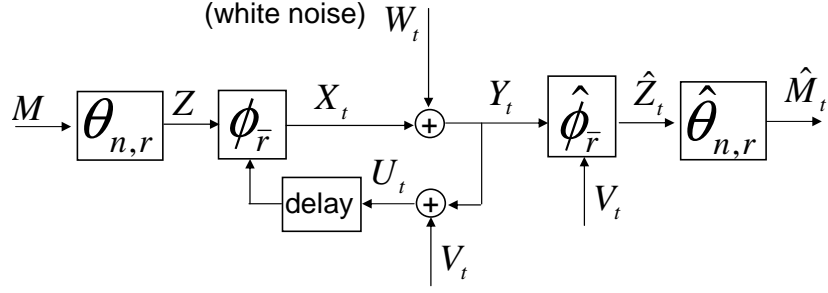,height=1.5 in,angle=0}

\caption{Basic feedback scheme with encoding and decoding.}
\label{Fig:1012pm31may06}
\end{figure}

In this Section, we describe a coding scheme in the presence of feedback according to the framework of Fig \ref{Fig:1012pm31may06}, where $\phi_{\bar{r}}$ and $\hat{\phi}_{\bar{r}}$ are defined by (\ref{eq:701pm31may06})-(\ref{eq:702pm31may06}), while the maps $\theta_{n,r}$ and $\hat{\theta}_{n,r}$ will be defined below. Notice that the scheme of Fig \ref{Fig:1012pm31may06} assumes that $\hat{\phi}_{\bar{r}}$ has
direct access to the feedback noise $V_t$. Under such an assumption, in this Section we construct
an efficient and simple coding and decoding scheme which will be used as a basic building block in the rest of the paper. 
In Section \ref{sec:447pm15jun06} we use the fact that if the backward link is corrupted by uniform quantization then, in fact, $V_t$ is the quantization error which can be recovered from the output of the forward channel and used as an input to $\hat{\phi}_{\bar{r}}$. Finally, in Section
\ref{sec:639pm17jun06} we show that bounded noise in the feedback
link can be dealt with by using a modification of the quantized feedback framework of
Section \ref{sec:447pm15jun06}. It should be noted that in the schemes presented in Sections \ref{sec:447pm15jun06} and \ref{sec:639pm17jun06}, the decoder relies solely on the output of the forward channel.

The main result of this Section is stated in Theorem \ref{th:313pm03may06}, where we
compute a rate of reliable\footnote{By reliable transmission we mean that the probability
of error converges to zero with increasing block length $n$.} transmission, in bits per channel use, which is achievable by 
the scheme of  Fig \ref{Fig:1012pm31may06}, in the presence of a
power constraint at the input of the forward channel. Such a transmission
rate is a function of the parameters $\sigma_W^2$,
$\bar{\sigma}_V$ and it also depends on the forward channel's input power
constraint, which we denote as $P_X^2$. Theorem \ref{th:313pm03may06} also
provides a lower bound on the error exponent of the resulting
scheme. If the forward channel is additive, white and Gaussian then
Theorem \ref{th:313pm03may06} shows that the probability of error
of the scheme of Fig \ref{Fig:1012pm31may06} decreases as a
doubly exponential function of the block length.

We start with the following definitions of the \emph{ceiling} and \emph{floor} functions denoted by $\bar{\Theta}$ and $\Theta$, respectively.
\begin{equation}
\bar{\Theta}(a)\overset{def}{=} \min \{n \in \mathbb{N}: a \leq n \}, \text{ } a \in \mathbb{R}
\end{equation}
\begin{equation}
\label{eq:353pm08june06}
\Theta(a) \overset{def}{=} \max \{n \in \mathbb{N}: a \geq n \}, \text{ } a \in \mathbb{R}
\end{equation}

The following definition specifies the maps $\theta_{n,r}$ and $\hat{\theta}_{n,r}$ represented in Fig \ref{Fig:1012pm31may06}.

\begin{definition} Given a positive integer $n$, a positive real constant $r$,
a random variable $M$ taking values in the set $\{1, \ldots,
2^{\Theta(rn)}\}$ and a real stochastic process $\hat{Z}_t$, the
following is the definition of the maps $\theta_{n,r}:M \mapsto Z$
and $\hat{\theta}_{n,r}:\hat{Z}_t \mapsto \hat{M}_t$:
\begin{equation}
\label{eq:1028pm31may06} Z= \theta_{n,r} (M) \overset{def}{=} \left(M-\frac{1}{2}
\right) 2^{-\Theta(rn)}
\end{equation}
\begin{equation}
\label{eq:1029pm31may06}  \hat{M}_t = \hat{\theta}_{n,r} (\hat{Z}_t)  \overset{def}{=}
\bar{\Theta}\left( 2^{\Theta(rt)} \hat{Z}_t \right), \text{ } t \in \{0, \ldots,n \}
\end{equation}
\end{definition}

For the remainder of this paper, $n$ denotes the block length of the coding schemes and $r$ represents a design parameter that quantifies the desired information rate, in bits per channel use. The following equations, describing the coding scheme of Fig \ref{Fig:1012pm31may06}, will be used in the statement of Lemma \ref{lem:948pm03may06} and Theorem \ref{th:313pm03may06}.

\begin{equation}
\label{eq:1129pm6sept06}
 \hat{M}_t=\hat{\theta}_{n,r} \left(\hat{\phi}_{\bar{r}}(t,Y^t,V^t)\right)
\end{equation}
\begin{equation}
Y_t=W_t+\underbrace{\phi_{\bar{r}} \left( t,\theta_{n,r}(M),U^{t-1} \right)}_{X_t}
\end{equation}
\begin{equation}
\label{eq:1130pm6sept06}
U_t=Y_t+V_t
\end{equation}

\begin{lemma}
\label{lem:948pm03may06}
Let $\sigma_W^2$, $\bar{\sigma}_V$ and $\bar{r}$ be given positive real parameters. Consider that the block length is given by a positive integer $n$, that the desired transmission rate is a positive real number  $r$ strictly less than $\bar{r}$ and that $M$ is a random variable arbitrarily distributed in the set $\{1,\ldots,2^{\Theta(rn)}\}$. If we adopt the scheme of Fig \ref{Fig:1012pm31may06}, alternatively described by (\ref{eq:1129pm6sept06})-(\ref{eq:1130pm6sept06}), then the following holds:
\begin{equation}
\label{eq:1136pm31may06}
Prob \left(M \neq \hat{M}_n \right) \leq \frac{ 2^{-2(\bar{r}-r)n }  E[X_n^2] }{4 (2^{\bar{r}}-2^{-\bar{r}})^{2}}
\end{equation} If $W_t$ is zero mean, white and Gaussian with variance $\sigma_W^2$ then the following doubly exponential decay, with increasing block size $n$, of the
probability of error holds:

\begin{equation}
\label{eq:517pm4sept06}
 Prob \left(M \neq \hat{M}_n \right) \leq e^{-\frac{1}{2 \beta^2}  \left( 2 (2^{\bar{r}}-2^{-\bar{r}}) 2^{(\bar{r}-r)n} -  \gamma\right)^2}
\end{equation} where $\gamma$ and $\beta$ are positive real constants given by (\ref{eq:258pm4sept06}) and (\ref{eq:259pm4sept06}), respectively.

\end{lemma}

\textbf{Proof:}
We start by using (\ref{eq:1028pm31may06})-(\ref{eq:1029pm31may06}) and the fact that $2^{\Theta(rn)}Z$ is in the set $ \{\frac{1}{2}, \ldots,2^{\Theta(rn)}-\frac{1}{2} \}$ to conclude the following:
\begin{equation}
\left| 2^{\Theta(rn)}Z - 2^{\Theta(rn)}\hat{Z}_n \right| < \frac{1}{2} \implies M=\hat{M}_n
\end{equation}
leading to:
\begin{equation}
\label{eq:1137pm31may06}
Prob\left(M \neq \hat{M}_n \right) \leq Prob \left( \left|Z -\hat{Z}_n \right| \geq  2^{-(\Theta(rn)+1)} \right)
\end{equation}

Using (\ref{eq:932pm31may06}), (\ref{eq:1137pm31may06}) and the fact that $\Theta(rn) \leq rn$, we get:

\begin{equation}
\label{eq:505pm4sept06}
Prob\left(M \neq \hat{M}_n \right) \leq Prob \left( |X_n| \geq 2 (2^{\bar{r}}-2^{-\bar{r}}) 2^{(\bar{r}-r)n} \right)
\end{equation}

The inequality (\ref{eq:1136pm31may06}) follows from  Markov's inequality applied to (\ref{eq:505pm4sept06}). Finally, the inequality (\ref{eq:517pm4sept06}) 
follows from (\ref{eq:505pm4sept06}) and (\ref{eq:257pm4sept06}). $\square$

\subsection{Lower-bounds on the achievable rate of reliable transmission in the presence of a power constraint at the input of the forward channel}
Below, we define a function that quantifies an achievable rate of reliable transmission for the scheme of Fig \ref{Fig:1012pm31may06}, in the presence of a power constraint at the input of the forward channel.
\begin{definition} 
\label{defvarrho}
For every choice of positive real parameters $\sigma_W^2$, $P_X^2$
and $\bar{\sigma}_V$ satisfying $ 4 \bar{\sigma}_V^2 \leq P_X^2$,
define a function $\varrho:(\sigma_W^2,P_X^2,\bar{\sigma}_V) \mapsto \mathbb{R}_{\geq 0}$ as
the non-negative real solution $\varrho$ of the following equation:
\begin{equation}
\label{eq:249pm03june06}
 \sigma_W \sqrt{2^{2\varrho}-1}=P_X-\bar{\sigma}_V \left( 1+2^{\varrho } \right)
\end{equation} If, instead, $ 4 \bar{\sigma}_V^2 > P_X^2$ then $\varrho(\sigma_W^2,P_X^2,\bar{\sigma}_V)\overset{\Delta}{=}0$.
\end{definition}
It is readily verifiable that a non-negative real solution of
(\ref{eq:249pm03june06}), in terms of
$\varrho$, exists and is unique,
provided that $\sigma_W^2$ and $P_X^2$ are strictly positive and
that $4 \bar{\sigma}_V^2$ is less or equal than $P_X^2$.

\begin{theorem}
\label{th:313pm03may06} Let $\sigma_W^2$, $P_X^2$ and
$\bar{\sigma}_V$  be given positive real parameters satisfying $4\bar{\sigma}_V^2 < P_X^2$. In
addition, select a positive transmission rate $r$ and a positive real constant $\bar{r}$
satisfying $r<\bar{r}<\varrho(\sigma_W^2,P_X^2,\bar{\sigma}_V)$. For every positive
integer block length $n$ the coding scheme of Fig \ref{Fig:1012pm31may06}, alternatively described by (\ref{eq:1129pm6sept06})-(\ref{eq:1130pm6sept06}),
leads to:

\begin{equation}
\label{eq:832pm16jun06}
E[X_t^2]\leq \left( P_X +  \underbrace{2^{-\bar{r}t} (2^{\bar{r}}-2^{-\bar{r}})  }_{\text{vanishes with increasing $t$}} \right)^2, 0 \leq t \leq n
\end{equation}

\begin{equation}
\label{eq:931pm4sept06}
Prob \left(M \neq \hat{M}_n \right) \leq \frac{ 2^{-2  \left( \bar{r} - r \right) n  }  E[X_n^2]}{ 4 (2^{\bar{r}}-2^{-\bar{r}})^{2} }
\end{equation} where $M$ is a random variable arbitrarily distributed in the set $\{1,\ldots,2^{\Theta(nr)}\}$. If $W_t$ is zero mean, white and Gaussian 
with variance $\sigma_W^2$ then the following doubly exponential decay, with increasing block size $n$, of the
probability of error holds:

\begin{equation}
\label{eq:902pm4sept06}
 Prob \left(M \neq \hat{M}_n \right) \leq e^{-\frac{1}{2 \beta^2}  \left( 2 (2^{\bar{r}}-2^{-\bar{r}}) 2^{(\bar{r}-r)n}  -  \gamma \right)^2}, \text{ } \alpha > 0
\end{equation} where $\gamma$ and $\beta$ are positive real constants given by (\ref{eq:258pm4sept06}) and (\ref{eq:259pm4sept06}), respectively.
\end{theorem}

Theorem \ref{th:313pm03may06} shows that the scheme of Fig
\ref{Fig:1012pm31may06}, under the constraint that the time average
of the second moment of $X_t$ is less or equal\footnote{See inequality (\ref{eq:832pm16jun06}).} than $P_X^2$, allows for reliable transmission
at any rate $r$ strictly less than $\varrho(\sigma_W^2,P_X^2,\bar{\sigma}_V)$. In addition, 
Theorem \ref{th:313pm03may06} shows that any rate of transmission $r$, if strictly less
than $\varrho(\sigma_W^2,P_X^2,\bar{\sigma}_V)$, leads to an
achievable error exponent arbitrarily close to
$2\left[r-\varrho(\sigma_W^2,P_X^2,\bar{\sigma}_V)\right]$. In addition, Theorem \ref{th:313pm03may06}
shows that if the forward channel is additive, white and Gaussian then the probability
of error decreases with the block length $n$ at a doubly exponential rate (see (\ref{eq:902pm4sept06})). 

\textbf{Proof of Theorem \ref{th:313pm03may06}:} The inequalities (\ref{eq:931pm4sept06}) and (\ref{eq:902pm4sept06}) follow directly from Lemma
\ref{lem:948pm03may06}. The derivation of (\ref{eq:832pm16jun06}) follows from (\ref{eq:1019pm30may06}) and from the fact that, from Definition \ref{defvarrho},
$\bar{r} < \varrho(\sigma_W^2,P_X^2,\bar{\sigma}_V)$ implies that $\sigma_W \sqrt{2^{2\bar{r}}-1}+\bar{\sigma}_V(2^{\bar{r}}+1) < P_X$. $\square$

\begin{figure}
\centering

\epsfig{file=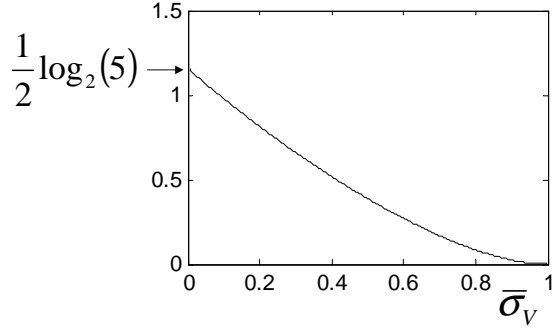,height=1.7 in,angle=0}

\caption{Plot of $\varrho(\sigma_W^2,P_X^2,\bar{\sigma}_V)$ using $\sigma_W^2=1$, $P_X^2=4$ and $\bar{\sigma}_V \in [0,1]$.}
\label{Fig:1112pm3june06}
\end{figure}

It follows from its definition, as the solution to
(\ref{eq:249pm03june06}), that
$\varrho(\sigma_W^2,P_X^2,\bar{\sigma}_V)$ also  satisfies the
following 3 properties:
\begin{equation}
\label{eq:424pm03june06}
\lim_{\bar{\sigma}_V \rightarrow 0^+}
\varrho(\sigma_W^2,P_X^2,\bar{\sigma}_V) = \frac{1}{2} \log_2 \left(1+\frac{P_X^2}{\sigma_W^2}\right),
\text{ } \sigma_W^2>0, \text{ } P_X^2 > 0
\end{equation}

\begin{equation}
\label{eq:1059pm06may06}
\varrho\left(\sigma_W^2,P_X^2,\frac{P_X^2}{4}\right) =0, \text{ } \sigma_W^2>0, \text{ } P_X^2 > 0
\end{equation}
\begin{equation} \label{eq: large P behavior}
\varrho(\sigma_W^2,P_X^2,\bar{\sigma}_V) \simeq \log_2
\left(\frac{P_X}{\sigma_W+\bar{\sigma}_V}\right), \text{ }  P_X^2 >>
\max \{\sigma_W^2,\bar{\sigma}_V^2\},
\end{equation}
where $\simeq$ indicates that the ratio between the left and right
hand sides of (\ref{eq: large P behavior}) tends to $1$ as $P_X
\rightarrow \infty$. If $W_t$ is white Gaussian then (\ref{eq:424pm03june06}) indicates that in
the limit, as the second moment of feedback noise goes to zero, the
scheme of Fig \ref{Fig:1012pm31may06} approaches capacity\footnote{It is a standard fact \cite{Thomas} that the capacity in bits per channel use of an additive Gaussian channel, with noise variance $\sigma_W^2$ and input power constraint $P_X^2$, is given by $\frac{1}{2} \log_2\left(1+\frac{P_X^2}{\sigma_W^2}\right)$.}. We have
computed $\varrho(\sigma_W^2,P_X^2,\bar{\sigma}_V)$ for
$\sigma_W^2=1$, $P_X^2=4$ and one thousand equally spaced values of
$\bar{\sigma}_V$, ranging from zero to one and the results are
plotted in Fig \ref{Fig:1112pm3june06}. The plot illustrates a graceful
(continuous) degradation of $\varrho(1,4,\bar{\sigma}_V)$ as a
function of $\bar{\sigma}_V$, going from the highest rate of
$\frac{1}{2}\log_2 5$, achieving capacity when $W_t$ is Gaussian,
down to zero when $\bar{\sigma}_V=1$, which is consistent with
(\ref{eq:424pm03june06}) and (\ref{eq:1059pm06may06}), respectively.

\section{Specification of a coding scheme using uniformly quantized feedback}
\label{sec:447pm15jun06}
\begin{figure}
\centering

\epsfig{file=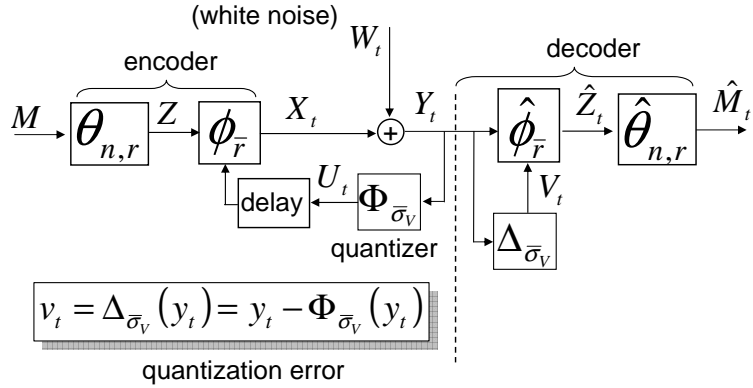,height=2 in,angle=0}

\caption{A coding and decoding scheme in the presence of uniformly quantized feedback.}
\label{Fig:658pm07june06}
\end{figure}

In this Section, we consider the scheme of Fig~\ref{Fig:658pm07june06}, where $\Phi_{\bar{\sigma}_V}$ represents a
 memoryless uniform quantizer with sensitivity $\bar{\sigma}_V$ and
$\Delta_{\bar{\sigma}_V}$ gives the associated quantization error. The main result of this Section is
Corollary \ref{cor:140pm13jun06}, where we indicate that the results of Section \ref{sec:231pm05june06} hold in the
presence of uniformly quantized feedback.
Notice that the diagram of Fig~\ref{Fig:658pm07june06} follows from
Fig~\ref{Fig:1012pm31may06} by adopting $V_t$ as the quantization
error, which the decoder re-constructs by making use of $\Delta_{\bar{\sigma}_V}$ applied to
the output of the forward channel. The precise definitions of the uniform quantizer
$\Phi_{\bar{\sigma}_V}$ and of the quantization error function
$\Delta_{\bar{\sigma}_V}$ are given below:

\begin{definition} Given a positive real parameter $b$, a uniform quantizer with sensitivity $b$ is a function $\Phi_b:\mathbb{R} \rightarrow \mathbb{R}$ defined as:
\begin{equation}
\Phi_b(y)=2b \Theta\left( \frac{y+b}{2b}\right)
\end{equation} where $\Theta$ is the \emph{floor} function specified in (\ref{eq:353pm08june06}). Similarly, the quantization error is given by the following function:
\begin{equation}
\Delta_b(y)=\Phi_b(y) - y, y \in \mathbb{R}
\end{equation} which satisfies the following bound:
\begin{equation}
\label{eq:216pm13june06}
| \Delta_b(y) | \leq b, y \in \mathbb{R}
\end{equation}
\end{definition}

The coding scheme of Fig \ref{Fig:658pm07june06} can be equivalently expressed by the following equations\footnote{Some of these equations have been used before, but we repeat them here for convenience.}:
\begin{equation}
\label{eq:1220pm7sept06}
\hat{M}_t=\hat{\theta}_{n,r} \left( \hat{\phi}_{\bar{r}}(t,Y^t,V^t) \right)
\end{equation}
\begin{equation}
Y_t=W_t+\underbrace{\phi_{\bar{r}} \left( t,\theta_{n,r}(M),U^{t-1} \right)}_{X_t}
\end{equation}
\begin{equation}
U_t=\Phi_{\bar{\sigma}_V}(Y_t)=Y_t+V_t
\end{equation}
\begin{equation}
\label{eq:1221pm7sept06}
V_t=\Delta_{\bar{\sigma}_V}(Y_t)
\end{equation}

The Corollary below follows directly from Theorem \ref{th:313pm03may06} applied to the scheme of Fig \ref{Fig:658pm07june06}, along with the upper-bound (\ref{eq:216pm13june06}).

\begin{corollary}
\label{cor:140pm13jun06}
Let $\sigma_W^2$, $P_X^2$ and $\bar{\sigma}_V$ be positive real constants satisfying $4\bar{\sigma}_V^2 < P_X^2$, where $\bar{\sigma}_V$ represents the sensitivity of the quantizer. In addition, select a positive transmission rate $r$ and a positive real constant $\bar{r}$ satisfying $r<\bar{r}<\varrho(\sigma_W^2,P_X^2,\bar{\sigma}_V^2)$. For every positive integer block length $n$, the coding scheme specified by (\ref{eq:1220pm7sept06})-(\ref{eq:1221pm7sept06}) (see Fig \ref{Fig:658pm07june06}) leads to:

\begin{equation}
 E[X_t^2] \leq \left( P_X + \underbrace{ 2^{-\bar{r}t} (2^{\bar{r}}-2^{-\bar{r}}) }_{\text{vanishes with increasing $t$}} \right)^2, 0 \leq t \leq n
\end{equation}
\begin{equation}
Prob \left(M \neq \hat{M}_n \right) \leq  \frac{  2^{- 2  \left( \bar{r} - r \right) n  }  E[X_n^2]}{4 (2^{\bar{r}}-2^{-\bar{r}})^{2} }
\end{equation} where $M$ is a random variable arbitrarily distributed in the set $\{1,\ldots,2^{\Theta(nr)}\}$. If $W_t$ is zero mean, white and Gaussian with variance $\sigma_W^2$
 then the following doubly exponential decay, with increasing block size $n$, of the
probability of error holds:
\begin{equation}
\label{eq:903pm4sept06}
 Prob \left(M \neq \hat{M}_n \right) \leq e^{-\frac{1}{2 \beta^2}  \left(2 (2^{\bar{r}}-2^{-\bar{r}}) 2^{(\bar{r}-r)n}  -  \gamma\right)^2}
\end{equation} where $\gamma$ and $\beta$ are positive real constants given by (\ref{eq:258pm4sept06}) and (\ref{eq:259pm4sept06}), respectively.
\end{corollary}

Notice that Corollary \ref{cor:140pm13jun06} shows that, in the presence of uniformly quantized feedback with sensitivity $\bar{\sigma}_V$, any rate $r$ strictly less than $\varrho(\sigma_W^2,P_X^2,\bar{\sigma}_V)$ allows for reliable transmission. This implies that the properties (\ref{eq:424pm03june06})-(\ref{eq:1059pm06may06}), along with the conclusions derived in Section \ref{sec:231pm05june06}, hold for uniformly quantized feedback. In particular, the achievable rate of reliable transmission of the coding scheme of Fig \ref{Fig:658pm07june06} degrades gracefully as a continuous function of the quantizer sensitivity $\bar{\sigma}_V$  (see the numerical example portrayed in Fig \ref{Fig:1112pm3june06}).

\section{Coding and decoding in the presence of feedback corrupted by bounded noise.}
\label{sec:639pm17jun06}

\begin{figure}
\centering

\epsfig{file=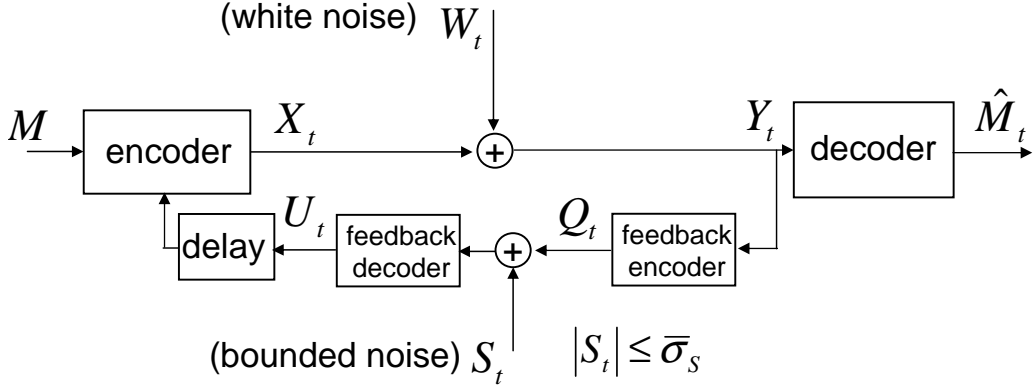,height=2 in,angle=0}

\caption{Communication scheme in the presence of bounded feedback noise.}
\label{Fig:124am16june06}
\end{figure}

From Corollary \ref{cor:140pm13jun06}, we conclude that there exist
simple explicit coding strategies based on Schalkwijk-Kailath's framework that, even in the
presence of \emph{uniformly quantized} feedback, provide positive rates with positive error exponents. 
 In this Section, 
we aim at designing coding schemes 
in the presence of feedback corrupted by bounded noise.
 The main result of
this Section is discussed in Section \ref{sec:712pm7jul06}, where we
describe a communication scheme whose structure
is that of Fig \ref{Fig:124am16june06}. In addition, we analyze the performance of
such a scheme in the presence of power constraints at the
input of the forward and backward channels. The proposed scheme
retains the simplicity of the Schalkwijk-Kailath scheme
\cite{SchalkwijkKailath}, but, in contrast to the original scheme
 (which breaks down in the presence of noise in the backward link
\cite[Section III.D]{SchalkwijkKailath}), achieves a positive rate
of reliable communication and is in fact capacity achieving in the
limit of high SNR in the backward link (assuming white Gaussian noise in
the forward channel). The scheme proposed in Section \ref{sec:712pm7jul06} 
also guarantees that, if the forward channel is additive, white and Gaussian, then the probability 
of error converges to zero as a doubly exponential function of the block length. The main results of
this Section are stated in Theorem \ref{thm:234pm5jul06}.

\subsection{Performance in the presence of a power constraint at the input of the backward channel.}
\label{sec:712pm7jul06}

For the remainder of this Section, we will define a coding scheme whose structure is that of Fig \ref{Fig:124am16june06}. The additive noise $S_t$ in the feedback link is 
arbitrarily distributed, bounded and the tightest upper-bound to its amplitude is defined below:$$\bar{\sigma}_S \overset{def}{=} \inf \{ \alpha \in \mathbb{R}_{\geq 0} : Prob(|S_t|>\alpha)=0, t \geq 0\}
$$ meaning that the following holds: $$Prob(|S_t| \leq \bar{\sigma}_S)=1, \text{ } t \geq 0$$ 

The following remark will be used in the construction of a coding scheme with the structure of Fig \ref{Fig:124am16june06}.

\begin{remark}
\label{rem:830pm15jun06}
Let $\bar{\sigma}_S$ be a positive real constant and $S_t$ be a real valued stochastic process satisfying $|S_t| \leq \bar{\sigma}_S$ with probability one. Given a positive real parameter $\bar{\sigma}_V$, the following holds with probability one:
\begin{equation}
\frac{\bar{\sigma}_V}{\bar{\sigma}_S} \Phi_{\bar{\sigma}_S} \left( S_t + Q_t \right) = \Phi_{\bar{\sigma}_V}(Y_t)
\end{equation} where $Q_t$ is given by:
\begin{equation}
\label{eq:1205pm16jun06}
Q_t=\Phi_{\bar{\sigma}_S} \left( \frac{\bar{\sigma}_S}{\bar{\sigma}_V} Y_t \right)
\end{equation}

\end{remark}

\begin{figure}
\centering

\epsfig{file=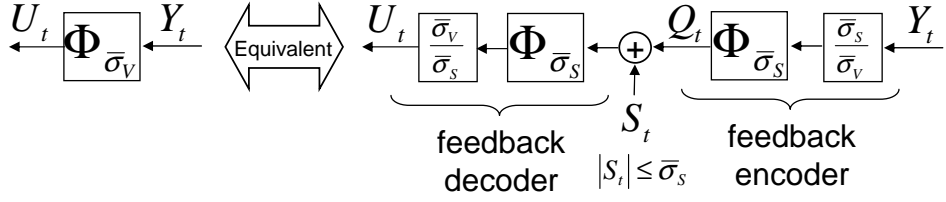,height=1 in,angle=0}

\caption{Schematic representation of the equivalence expressed in Remark \ref{rem:830pm15jun06}.}
\label{Fig:829pm15june06}
\end{figure}

The schematic representation of the equivalence expressed in Remark \ref{rem:830pm15jun06} is displayed in Fig~\ref{Fig:829pm15june06}. In such a scheme, $S_t$ is the bounded additive noise at the backward channel with input $Q_t$.

\begin{figure}
\centering

\epsfig{file=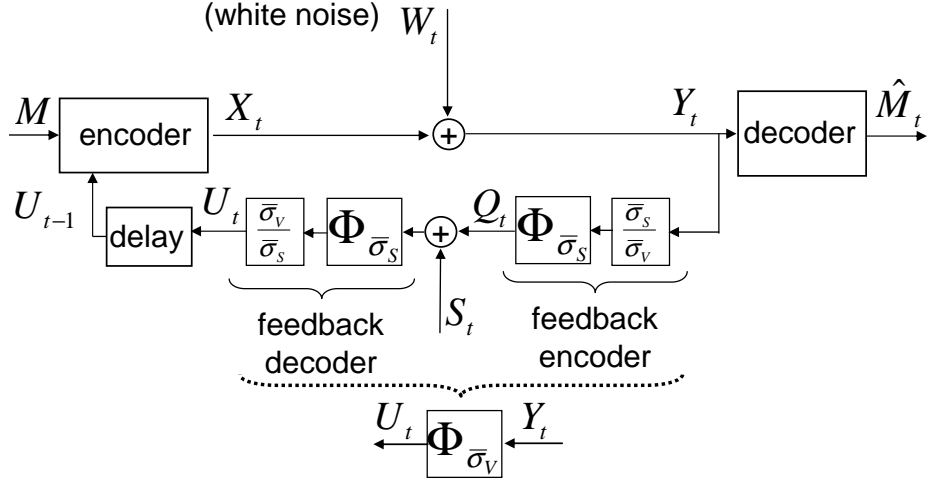,height=2.5 in,angle=0}

\caption{Proposed coding scheme for dealing with the feedback corruption that results from additive noise $S_t$ in the backward channel. The scheme is constructed by replacing the quantizer in the diagram of Fig \ref{Fig:658pm07june06} with the scheme of Fig \ref{Fig:829pm15june06}. The encoder and decoder blocks are described in detail in Fig \ref{Fig:658pm07june06}.}
\label{Fig:118pm17aug06}
\end{figure}

Aiming at constructing a coding scheme according to the structure of Fig \ref{Fig:124am16june06}, 
we use Remark \ref{rem:830pm15jun06} to obtain a new coding strategy by
substituting the feedback quantizer $\Phi_{\bar{\sigma}_V}$ of Fig
\ref{Fig:658pm07june06} with the equivalent additive noise channel
diagram of Fig \ref{Fig:829pm15june06}. The resulting scheme, along
with the encoding and decoding strategy of Section
\ref{sec:447pm15jun06}, provides a solution to the problem of
designing encoders and decoders in the presence of an additive
(bounded) noise backward channel (see Fig \ref{Fig:118pm17aug06}).
Under such a design strategy, $\bar{\sigma}_V$ becomes a design
parameter. Notice that viewing $\bar{\sigma}_V$ as a design knob is
in contrast with the framework of Section
\ref{sec:447pm15jun06}, where $\bar{\sigma}_V$ was a given constant.

Regarding the role of $\bar{\sigma}_V$, we have shown in (\ref{eq:424pm03june06}) that as $\bar{\sigma}_V$ approaches zero the achievable rate of reliable transmission converges to a positive value, which, in the case where $W_t$ is white Gaussian, coincides with capacity. However, for any given positive real $\bar{\sigma}_S$, the smaller $\bar{\sigma}_V$ the larger the scaling constant $\frac{\bar{\sigma}_S}{\bar{\sigma}_V}$ in (\ref{eq:1205pm16jun06}) and that may lead to $Q_t$ having an arbitrarily large second moment. In Theorem \ref{thm:234pm5jul06}, we show that the function defined below solves the aforementioned problem by providing a suitable choice for $\sigma_V$, in the presence of power constraints at the input of the forward and backward channels.

\begin{definition} Let $\sigma_W$, $\bar{\sigma}_S$, $P_X$ and $P_Q$ be given positive real constants, where $P_Q^2$ symbolizes a power constraint at the input of the backward channel $Q_t$. Below, we define the function $\Gamma:\mathbb{R}_{\geq 0}^4 \rightarrow \mathbb{R}_{\geq 0}$, which we will use as a selection for the design parameter $\bar{\sigma}_V$:
\begin{equation}
\Gamma\left(\sigma_W,\bar{\sigma}_S,P_X,P_Q \right)=
\left(P_X+\sigma_W \right)\frac{\bar{\sigma}_S}{P_Q-\bar{\sigma}_S}, \text{ } P_Q>\bar{\sigma}_S
\end{equation}
\end{definition}

The following Theorem is one of the main results of this paper.

\begin{theorem}
\label{thm:234pm5jul06}
Let $\sigma_W^2$, $P_X^2$, $P_Q^2$ and $\bar{\sigma}_S$ be positive constants satisfying $4\Gamma\left(\sigma_W,\bar{\sigma}_S,P_X,P_Q \right)^2 < P_X^2$ and $\bar{\sigma}_S < P_Q$. In addition, select a positive transmission rate $r$ and a positive real constant $\bar{r}$ satisfying $r<\bar{r}<\left. \varrho(\sigma_W^2,P_X^2,\bar{\sigma}_V) \right|_{\bar{\sigma}_V=\Gamma\left(\sigma_W,\bar{\sigma}_S,P_X,P_Q \right)}$.  For every positive integer block length $n$, the coding scheme of Fig \ref{Fig:118pm17aug06}, alternatively described by (\ref{eq:1220pm7sept06})-(\ref{eq:1221pm7sept06}) and (\ref{eq:1205pm16jun06}), leads to:

\begin{equation}
\label{eq:1237pm5jul06} E[X_t^2] \leq \left( P_X +
\underbrace{2^{-\bar{r}t} \left(
2^{\bar{r}}-2^{-\bar{r}}\right)}_{\text{vanishes with increasing
$t$}} \right)^2, \text{ }  0 \leq t \leq n
\end{equation}

\begin{equation}
\label{eq:1238pm5jul06}
  E[Q_t^2] \leq \left( P_Q+\underbrace{ 2^{-\bar{r} t} \frac{P_Q-\bar{\sigma}_S}{ P_X+\sigma_W}  \left( 2^{\bar{r}}-2^{-\bar{r}}\right)  }_{
  \text{vanishes with increasing $t$}} \right)^2, \text{ }  0 \leq t \leq n
\end{equation}

\begin{equation}
\label{eq:1239pm5jul06}
Prob \left(M \neq \hat{M}_n \right) \leq  \frac{ 2^{-2  \left( \bar{r} - r \right) n  } E[X_n^2] }{4  (2^{\bar{r}}-2^{-\bar{r}})^{2} }
\end{equation} where $M$ is a random variable arbitrarily distributed in the set $\{1,\ldots,2^{\Theta(nr)}\}$. If $W_t$ is zero mean, white and Gaussian with variance 
$\sigma_W^2$ then the following doubly exponential decay, with increasing block size $n$, of the
probability of error holds:

\begin{equation}
\label{eq:905pm4sept06}
 Prob \left(M \neq \hat{M}_n \right) \leq e^{-\frac{1}{2 \beta^2}  \left( 2 (2^{\bar{r}}-2^{-\bar{r}}) 2^{(\bar{r}-r)n} -  \gamma \right)^2}
\end{equation} where $\gamma$ and $\beta$ are positive real constants given by (\ref{eq:258pm4sept06}) and (\ref{eq:259pm4sept06}), respectively,
where $\bar{\sigma}_V $ is given by the assumed selection  $\bar{\sigma}_V = \Gamma\left(\sigma_W,\bar{\sigma}_S,P_X,P_Q \right)$.
\end{theorem}

\textbf{Proof:}
The inequalities (\ref{eq:1237pm5jul06}), (\ref{eq:1239pm5jul06}) and (\ref{eq:905pm4sept06}) follow directly from Corollary \ref{cor:140pm13jun06}. 
In order to arrive at (\ref{eq:1238pm5jul06}), we start by noticing that we can use the triangular inequality to find the following inequalities:
\begin{equation}
\label{eq:808pm16jun06}
\left( E[Y_t^2] \right)^{\frac{1}{2}} \leq \left( E[X_t^2]\right)^{\frac{1}{2}} + \sigma_W
\end{equation}

\begin{equation}
\label{eq:809pm16jun06}
\left( E[Q_t^2] \right)^{\frac{1}{2}} \leq \frac{\bar{\sigma}_S}{\bar{\sigma}_V}\left( E[Y_t^2]\right)^{\frac{1}{2}} + \bar{\sigma}_S
\end{equation}

In addition, substitution of (\ref{eq:808pm16jun06}) in (\ref{eq:809pm16jun06}), leads to:

\begin{equation}
\label{eq:802pm16jun06}
E[Q_t^2] \leq \left( \frac{\bar{\sigma}_S}{\bar{\sigma}_V} \left( E[X_t^2]^{\frac{1}{2}}+\sigma_W\right)+\bar{\sigma}_S \right)^2
\end{equation} which, from (\ref{eq:1237pm5jul06}), implies the following:
\begin{equation}
\label{eq:632pm5sept06}
E[Q_t^2] \leq \left( \frac{\bar{\sigma}_S}{\bar{\sigma}_V} \left( P_X +
2^{-\bar{r}t} \left(2^{\bar{r}}-2^{-\bar{r}}\right)+\sigma_W\right)+\bar{\sigma}_S \right)^2
\end{equation}

The proof is complete since (\ref{eq:1238pm5jul06}) follows by substituting our choice $\bar{\sigma}_V=\Gamma\left(\sigma_W,\bar{\sigma}_S,P_X,P_Q \right)$ in (\ref{eq:632pm5sept06}). $\square$

Under the conditions of Theorem \ref{thm:234pm5jul06}, including our choice of the design parameter $\bar{\sigma}_V$, the following limit holds:

\begin{equation}
\label{eq:651pm17jun06}
\lim_{\bar{\sigma}_S \rightarrow 0^+} \left. \varrho(\sigma_W^2,P_X^2,\bar{\sigma}_V)\right|_{\bar{\sigma}_V=\Gamma\left(\sigma_W,\bar{\sigma}_S,P_X,P_Q \right)} = \frac{1}{2} \log_2 \left( 1+ \frac{P_X^2}{\sigma_W^2} \right), \text{ } \sigma_W>0, P_X >0, P_Q>0
\end{equation}

\begin{figure}
\centering

\epsfig{file=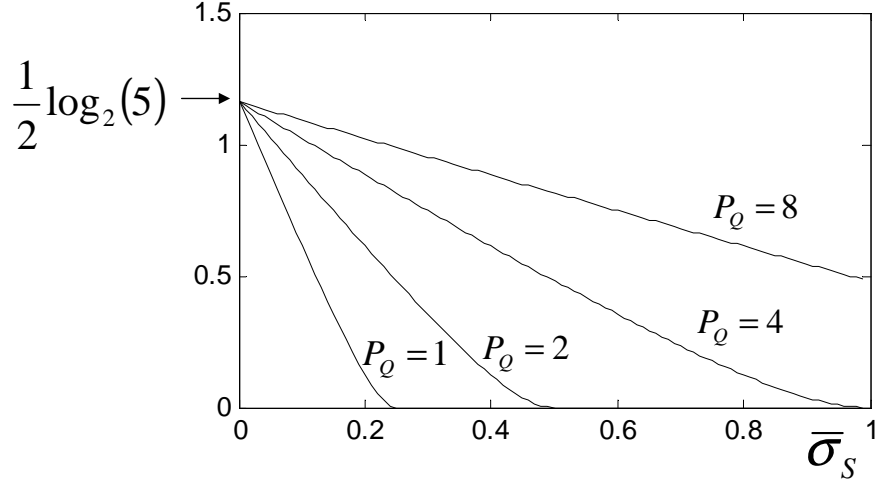,height=2.5 in,angle=0}

\caption{Plot of $ \left. \varrho(\sigma_W^2,P_X^2,\bar{\sigma}_V)\right|_{\bar{\sigma}_V=\Gamma\left(\sigma_W,\bar{\sigma}_S,P_X,P_Q \right)}$ using $\sigma_W^2=1$, $P_X^2=4$ and $\bar{\sigma}_S \in [0,1)$, for $P_Q$ taking values $1$, $2$, $4$ and $8$.}
\label{Fig:315pm5jul06}
\end{figure}

Notice that (\ref{eq:651pm17jun06}) leads to the conclusion that,
under our choice of $\bar{\sigma}_V$, the performance of
the scheme of Theorem \ref{thm:234pm5jul06} (see Fig
\ref{Fig:118pm17aug06}) degrades gracefully as a function of
$\bar{\sigma}_S$, in terms of both the rate and the error exponent.
If $W_t$ is white Gaussian then (\ref{eq:651pm17jun06}) indicates that as
$\bar{\sigma}_S$ tends to zero, the scheme of Theorem
\ref{thm:234pm5jul06} can be used to reliably communicate at a rate
arbitrarily close to capacity. Moreover, such a conclusion holds in
the presence of an arbitrarily low power constraint at the backward
channel. The plot of Fig \ref{Fig:315pm5jul06} displays how the
achievable rate changes as a function of $\bar{\sigma}_S$, under the
choice $\bar{\sigma}_V=\Gamma\left(\sigma_W,\bar{\sigma}_S,P_X,P_Q
\right)$. Such a plot also illustrates that by increasing $P_Q$ we
can reduce the sensitivity of the achievable rate, of reliable transmission, 
relative to variations in $\bar{\sigma}_S$.

\subsection{Further comments on the location of the one-step feedback delay}

\begin{figure}
\centering

\epsfig{file=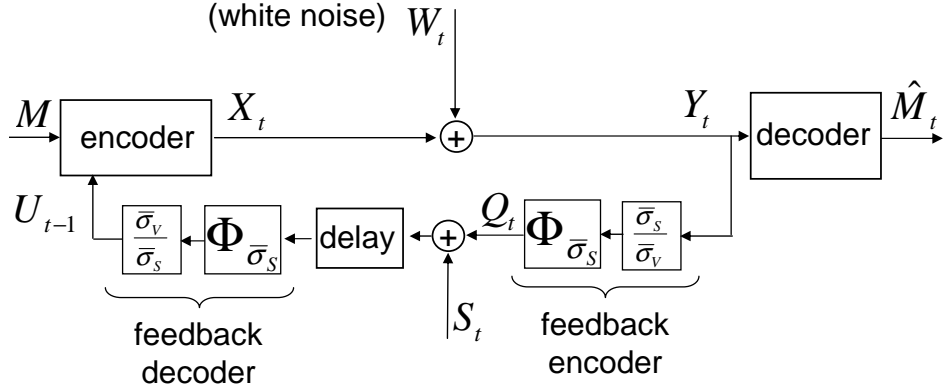,height=2 in,angle=0}

\caption{A coding scheme equivalent to the one described by Fig \ref{Fig:118pm17aug06}. }
\label{Fig:742pm5sept06}
\end{figure}

In the framework of Fig \ref{Fig:118pm17aug06}, the one-step delay block is located \emph{after} the feedback decoder. However, we should stress that, since the feedback decoder is
time-invariant, our coding scheme would be unaltered if we had placed the delay block \emph{before} as indicated in  Fig \ref{Fig:742pm5sept06}. Indeed, the
diagrams of Fig \ref{Fig:118pm17aug06} and \ref{Fig:742pm5sept06} are equivalent, implying that Theorem \ref{thm:234pm5jul06} holds also for the coding
scheme of Fig \ref{Fig:742pm5sept06}.

\section{Conclusions}
\label{sec:conclusions} We derived simple schemes for reliable
communication over a white noise forward channel, in the presence 
of corrupted feedback. Both the case of uniform quantization noise and the case of
additive bounded noise in the backward link were considered, where, in the latter case,
 encoding at the input to the
backward channel is allowed. The schemes were seen to achieve a
positive rate of reliable communication, and in fact be
capacity-achieving in the presence of an additive white Gaussian forward channel,
in the limit of small noise (or high SNR when encoding is allowed)
in the backward link. In addition, still under the assumption that
the forward channel is additive white Gaussian, the proposed schemes guarantee
that the probability of error converges to zero as a doubly
exponential function of the block length.

We believe that our approach to the construction and analysis of coding schemes
carries over naturally to the case where the noise in the forward
channel is non-white. In this case, we expect to obtain variations on the
schemes in \cite{Kim} that are analogous to those in the present
work and whose gap to capacity behaves similarly.



\end{document}